\documentclass[12pt]{article}

\usepackage[body={17.5cm, 21cm},right=2cm]{geometry}

\usepackage{color}
\usepackage{graphicx}
\usepackage{epsf}
\usepackage{graphicx,epsfig}
\usepackage{amsmath}
\usepackage{amssymb}
\usepackage{epsfig}
\usepackage{cite}
\usepackage{color,colordvi}
\newcommand{\be}{\begin{eqnarray}}
\newcommand{\ee}{\end{eqnarray}}
\newcommand{\bi}{\begin{itemize}}
\newcommand{\ei}{\end{itemize}}

\newcommand{\bx}{{\bf{x}}}

\newcounter{hran}

\def\MSbar{\relax\ifmmode\overline{\rm MS}\else{$\overline{\rm MS}${ }}\fi}

\def\d{\rm d}

 \def\vx{\vec{ x}} 
\def\vk{\vec{k}}

\def\vx{\vec{x}}

\def\bx{{\vec{x}}}
\def\bk{{\vec{k}}}

\def\bv{{\vec{v}}}
\def\t{\tau}
\def\ba{{\vec{g}}}

\def\tr{{\rm tr}}
\def\d{\rm d}

\def\simlt{\stackrel{<}{{}_\sim}}

\numberwithin{equation}{section}
\begin{document}\thispagestyle{empty}
\vspace{5mm}
\vspace{0.5cm}
\begin{center}

\def\thefootnote{\fnsymbol{footnote}}

{\large \bf 
  Conformal Symmetries of FRW Accelerating Cosmologies 
}
\\[1.5cm]
{\large  A. Kehagias$^{a,b}$ and A. Riotto$^{b}$}
\\[0.5cm]

\vspace{.3cm}
{\normalsize {\it  $^{a}$ Physics Division, National Technical University of Athens, \\15780 Zografou Campus, Athens, Greece}}\\

\vspace{.3cm}
{\normalsize { \it $^{b}$ Department of Theoretical Physics and Center for Astroparticle Physics (CAP)\\ 24 quai E. Ansermet, CH-1211 Geneva 4, Switzerland}}\\

\vspace{.3cm}


\end{center}

\vspace{3cm}

\hrule \vspace{0.3cm}
{\small  \noindent \textbf{Abstract} \\[0.3cm]
\noindent 
We show that  any  accelerating Friedmann-Robertson-Walker (FRW) cosmology
with equation of state $w<-1/3$ (and therefore not only a de Sitter stage with $w=-1$) exhibits three-dimensional conformal symmetry on
 future constant-time hypersurfaces.   We also offer an alternative derivation of this result in terms of conformal Killing vectors  and show that 
 long wavelength comoving curvature perturbations of the perturbed FRW metric are just conformal Killing motions of the FRW background.
 We then  extend theb boundary  conformal symmetry to the  bulk for accelerating cosmologies.
Our findings indicate  that  one can easily generate perturbations of scalar fields which are  not only scale invariant, but also fully conformally invariant on super-Hubble scales. Measuring  a scale-invariant power spectrum for the cosmological perturbation does not automatically imply that the universe went through a de Sitter stage.

\vspace{0.5cm}  \hrule
\vskip 1cm

\def\thefootnote{\arabic{footnote}}
\setcounter{footnote}{0}



\baselineskip= 19pt

\newpage 
\tableofcontents

\newcommand{\fix}{\Phi(\mathbf{x})}
\newcommand{\fiLx}{\Phi_{\rm L}(\mathbf{x})}
\newcommand{\fiNLx}{\Phi_{\rm NL}(\mathbf{x})}
\newcommand{\fik}{\Phi(\mathbf{k})}
\newcommand{\fiLk}{\Phi_{\rm L}(\mathbf{k})}
\newcommand{\fiLkone}{\Phi_{\rm L}(\mathbf{k_1})}
\newcommand{\fiLktwo}{\Phi_{\rm L}(\mathbf{k_2})}
\newcommand{\fiLkthree}{\Phi_{\rm L}(\mathbf{k_3})}
\newcommand{\fiLkfour}{\Phi_{\rm L}(\mathbf{k_4})}
\newcommand{\fiNLk}{\Phi_{\rm NL}(\mathbf{k})}
\newcommand{\fiNLkone}{\Phi_{\rm NL}(\mathbf{k_1})}
\newcommand{\fiNLktwo}{\Phi_{\rm NL}(\mathbf{k_2})}
\newcommand{\fiNLkthree}{\Phi_{\rm NL}(\mathbf{k_3})}

\newcommand{\kernel}{f_{\rm NL} (\mathbf{k_1},\mathbf{k_2},\mathbf{k_3})}
\newcommand{\dirac}{\delta^{(3)}\,(\mathbf{k_1+k_2-k})}
\newcommand{\dirackonektwokthree}{\delta^{(3)}\,(\mathbf{k_1+k_2+k_3})}

\newcommand{\beq}{\begin{equation}}
\newcommand{\eeq}{\end{equation}}
\newcommand{\beqarr}{\begin{eqnarray}}
\newcommand{\eeqarr}{\end{eqnarray}}

\newcommand{\angk}{\hat{k}}
\newcommand{\angn}{\hat{n}}

\newcommand{\tfnow}{\Delta_\ell(k,\tau_0)}
\newcommand{\tf}{\Delta_\ell(k)}
\newcommand{\tfone}{\Delta_{\el\ell_1}(k_1)}
\newcommand{\tftwo}{\Delta_{\el\ell_2}(k_2)}
\newcommand{\tfthree}{\Delta_{\el\ell_3}(k_3)}
\newcommand{\tffour}{\Delta_{\el\ell_1^\prime}(k)}
\newcommand{\deltatilde}{\widetilde{\Delta}_{\el\ell_3}(k_3)}

\newcommand{\alm}{a_{\ell m}}
\newcommand{\almL}{a_{\ell m}^{\rm L}}
\newcommand{\almNL}{a_{\ell m}^{\rm NL}}
\newcommand{\almone}{a_{\el\ell_1 m_1}}
\newcommand{\almLone}{a_{\el\ell_1 m_1}^{\rm L}}
\newcommand{\almNLone}{a_{\el\ell_1 m_1}^{\rm NL}}
\newcommand{\almtwo}{a_{\el\ell_2 m_2}}
\newcommand{\almLtwo}{a_{\el\ell_2 m_2}^{\rm L}}
\newcommand{\almNLtwo}{a_{\el\ell_2 m_2}^{\rm NL}}
\newcommand{\almthree}{a_{\el\ell_3 m_3}}
\newcommand{\almLthree}{a_{\el\ell_3 m_3}^{\rm L}}
\newcommand{\almNLthree}{a_{\el\ell_3 m_3}^{\rm NL}}

\newcommand{\YLMstar}{Y_{L M}^*}
\newcommand{\Ylmstar}{Y_{\ell m}^*}
\newcommand{\Ylmstarone}{Y_{\el\ell_1 m_1}^*}
\newcommand{\Ylmstartwo}{Y_{\el\ell_2 m_2}^*}
\newcommand{\Ylmstarthree}{Y_{\el\ell_3 m_3}^*}
\newcommand{\Ylmstarfour}{Y_{\el\ell_1^\prime m_1^\prime}^*}
\newcommand{\Ylmstarfive}{Y_{\el\ell_2^\prime m_2^\prime}^*}
\newcommand{\Ylmstarsix}{Y_{\el\ell_3^\prime m_3^\prime}^*}

\newcommand{\YLM}{Y_{L M}}
\newcommand{\Ylm}{Y_{\ell m}}
\newcommand{\Ylmone}{Y_{\el\ell_1 m_1}}
\newcommand{\Ylmtwo}{Y_{\el\ell_2 m_2}}
\newcommand{\Ylmthree}{Y_{\el\ell_3 m_3}}
\newcommand{\Ylmfour}{Y_{\el\ell_1^\prime m_1^\prime}}
\newcommand{\Ylmfive}{Y_{\el\ell_2^\prime m_2^\prime}}
\newcommand{\Ylmsix}{Y_{\el\ell_3^\prime m_3^\prime}}

\newcommand{\comm}[1]{\textbf{\textcolor{rossos}{#1}}}
\newcommand{\lsim}{\,\raisebox{-.1ex}{$_{\textstyle <}\atop^{\textstyle\sim}$}\,}
\newcommand{\gsim}{\,\raisebox{-.3ex}{$_{\textstyle >}\atop^{\textstyle\sim}$}\,}

\newcommand{\jl}{j_\ell(k r)}
\newcommand{\jlfourone}{j_{\el\ell_1^\prime}(k_1 r)}
\newcommand{\jlfivetwo}{j_{\el\ell_2^\prime}(k_2 r)}
\newcommand{\jlsixthree}{j_{\el\ell_3^\prime}(k_3 r)}
\newcommand{\jlsix}{j_{\el\ell_3^\prime}(k r)}
\newcommand{\jlthree}{j_{\el\ell_3}(k_3 r)}
\newcommand{\jlthreetau}{j_{\el\ell_3}(k r)}

\newcommand{\Gaunt}{\mathcal{G}_{\el\ell_1^\prime \, \el\ell_2^\prime \, 
\el\ell_3^\prime}^{m_1^\prime m_2^\prime m_3^\prime}}
\newcommand{\Gaunttwo}{\mathcal{G}_{\el\ell_1^\prime \, \el\ell_2^\prime \, 
\el\ell_3}^{m_1^\prime m_2^\prime m_3}}
\newcommand{\Gauntstardef}{\mathcal{H}_{\el\ell_1 \, \el\ell_2 \, \el\ell_3}^{m_1 m_2 m_3}}
\newcommand{\Gauntstarone}{\mathcal{G}_{\el\ell_1 \, L \,\, \el\ell_1^\prime}
^{-m_1 M m_1^\prime}}
\newcommand{\Gauntstartwo}{\mathcal{G}_{\el\ell_2^\prime \, \el\ell_2 \, L}
^{-m_2^\prime m_2 M}}

\newcommand{\de}{{\rm d}}

\newcommand{\dangn}{d \angn}
\newcommand{\dangk}{d \angk}
\newcommand{\dangkone}{d \angk_1}
\newcommand{\dangktwo}{d \angk_2}
\newcommand{\dangkthree}{d \angk_3}
\newcommand{\dk}{d^3 k}
\newcommand{\dkone}{d^3 k_1}
\newcommand{\dktwo}{d^3 k_2}
\newcommand{\dkthree}{d^3 k_3}
\newcommand{\dkfour}{d^3 k_4}
\newcommand{\dallk}{\dkone \dktwo \dk}

\newcommand{\FT}{ \int  \! \frac{d^3k}{(2\pi)^3} 
e^{i\mathbf{k} \cdot \angn \tau_0}}
\newcommand{\planewave}{e^{i\mathbf{k \cdot x}}}
\newcommand{\dallkfourier}{\frac{\dkone}{(2\pi)^3}\frac{\dktwo}{(2\pi)^3}
\frac{\dkthree}{(2\pi)^3}}

\newcommand{\Bis}{B_{\el\ell_1 \el\ell_2 \el\ell_3}^{m_1 m_2 m_3}}
\newcommand{\Avbis}{B_{\el\ell_1 \el\ell_2 \el\ell_3}}

\newcommand{\los}{\mathcal{L}_{\el\ell_3 \el\ell_1 \el\ell_2}^{L \, 
\el\ell_1^\prime \el\ell_2^\prime}(r)}
\newcommand{\loszero}{\mathcal{L}_{\el\ell_3 \el\ell_1 \el\ell_2}^{0 \, 
\el\ell_1^\prime \el\ell_2^\prime}(r)}
\newcommand{\losone}{\mathcal{L}_{\el\ell_3 \el\ell_1 \el\ell_2}^{1 \, 
\el\ell_1^\prime \el\ell_2^\prime}(r)}
\newcommand{\lostwo}{\mathcal{L}_{\el\ell_3 \el\ell_1 \el\ell_2}^{2 \, 
\el\ell_1^\prime \el\ell_2^\prime}(r)}
\newcommand{\losfNL}{\mathcal{L}_{\el\ell_3 \el\ell_1 \el\ell_2}^{0 \, 
\el\ell_1 \el\ell_2}(r)}



\def\d{\partial}
\def\C{{\rm CDM}}
\def\me{m_e}
\def\te{T_e}
\def\ti{\tau_{\rm initial}}
\def\tci#1{n_e(#1) \sigma_T a(#1)}
\def\tr{\tau_r}
\def\dtr{\delta\tau_r}
\def\dd{\widetilde\Delta^{\rm Doppler}}
\def\dsw{\Delta^{\rm Sachs-Wolfe}}
\def\clsw{C_\ell^{\rm Sachs-Wolfe}}
\def\cldop{C_\ell^{\rm Doppler}}
\def\Dt{\widetilde{\Delta}}
\def\mut{\mu}
\def\vt{\widetilde v}
\def\hp{ {\bf \hat p}}
\def\sdv{S_{\delta v}}
\def\svv{S_{vv}}
\def\bvt{\widetilde{\bv}}
\def\delt{\widetilde{\delta_e}}
\def\cos{{\rm cos}}
\def\nn{\nonumber \\}
\def\bq{ {\bf q} }
\def\ba{ {\bf p} }
\def\bap{ {\bf p'} }
\def\bqp{ {\bf q'} }
\def\bp{ {\bf p} }
\def\bpp{ {\bf p'} }
\def\bk{ {\bf k} }
\def\bx{ {\bf x} }
\def\bv{ {\bf v} }
\def\qp{ p^{\mu}k_{\mu} }
\def\qpp { p^{\mu} k'_{\mu} }
\def\bgm{ {\bf \gamma} }
\def\bkp{ {\bf k'} }
\def\gq{ g(\bq)}
\def\gqp{ g(\bqp)}
\def\fp{ f(\bp)}
\def\h#1{ {\bf \hat #1}}
\def\fpp{ f(\bpp)}
\def\fz{f^{(\vec{0})}(p)}
\def\fpz{f^{(\vec{0})}(p')}
\def\f#1{f^{(#1)}(\bp)}
\def\fps#1{f^{(#1)}(\bpp)}
\def\dq{ {d^3\bq \over (2\pi)^32E(\bq)} }
\def\dqp{ {d^3\bqp \over (2\pi)^32E(\bqp)} }
\def\dpp{ {d^3\bpp \over (2\pi)^32E(\bpp)} }
\def\dtq{ {d^3\bq \over (2\pi)^3} }
\def\dtqp{ {d^3\bqp \over (2\pi)^3} }
\def\dtpp{ {d^3\bpp \over (2\pi)^3} }
\def\part#1;#2 {\partial#1 \over \partial#2}
\def\deriv#1;#2 {d#1 \over d#2}
\def\Done{\Delta^{(1)}}
\def\Dtwo{\widetilde\Delta^{(2)}}
\def\fone{f^{(1)}}
\def\ftwo{f^{(2)}}
\def\tg{T_\gamma}
\def\delpp{\delta(p-p')}
\def\delb{\delta_B}
\def\tc{\tau_0}
\def\DD{\langle|\Delta(k,\mu,\tau_0)|^2\rangle}
\def\DDL{\langle|\Delta(k=l/\tc,\mu)|^2\rangle}
\def\bkpp{{\bf k''}}
\def\kmkp{|\bk-\bkp|}
\def\kmkpsq{k^2+k'^2-2kk'x}
\def\tt{ \left({\tau' \over \tau_c}\right)}
\def\kt{ k\mu \tau_c}

%
%
%

\section{Introduction}\pagenumbering{arabic}
\noindent
It has been appreciated for some time now that symmetries play a crucial role    in  characterizing  the properties of  the  cosmological perturbations  generated during a de Sitter  stage \cite{lrreview}.  Indeed,   the de Sitter isometry group acts  like  conformal group  on $\mathbb{R}^3$ when the fluctuations are on 
super-Hubble scales. As a consequence, the correlators of scalar fields, which are not the inflaton,  are constrained by conformal invariance as  the SO(1,4) isometry
of the de Sitter background is realized as conformal symmetry of the flat $\mathbb{R}^3$ sections  
\cite{antoniadis,maldacena1,creminelli1,us1,us2}.  
The  fact that the de Sitter isometry group acts as conformal group on the three-dimensional Euclidean space on super-Hubble scales can be also used to characterize the correlators involving the inflaton and  vector fields \cite{vec}.
In the  case in which the inflationary perturbations
originate from  the inflaton itself,   one can find conformal consistency relations among the inflationary correlators \cite{creminelli2,hui,baumann1,baumann2,mcfadden,mata,hui2}.   
Finally, consistency relations involving the soft limit of the $(n + 1)$-correlator functions of matter and galaxy overdensities have  also be found by investigating  the symmetries enjoyed by the Newtonian equations of motion of the non-relativistic dark matter fluid coupled to gravity
\cite{krls,ppls}. 

In this paper we wish to make the  simple, yet relevant, observation that  any  accelerating Friedmann-Robertson-Walker (FRW) cosmology
with equation of state $w<-1/3$ (and therefore not only a de Sitter stage with $w=-1$) exhibits three-dimensional conformal 
symmetry on super-Hubble scales.  This is because 
  for an accelerating universe the future   constant-time boundary possesses ten conformal Killing vectors  which form an 
  SO(1,4) algebra, 
  precisely the three-dimensional conformal algebra. Of course, only  for exact de Sitter space these ten 
  conformal Killing vectors are actually isometries,  whereas for all the other cases 
  only translations and rotations are isometries. 
  Our observation implies that if one constructs a theory invariant under translations, rotations, dilations and special conformal transformations
 (the generators of the three-dimensional conformal field group in  $\mathbb{R}^3$) and coupled it to 
 any accelerating FRW metric, then the full theory is 
  automatically SO(1,4) invariant and therefore conformal invariant at the future boundary. 
  This means that correlators of the super-Hubble fluctuations of a scalar field must satisfy such a 
  three-dimensional conformal field symmetry. By using the set of available conformal Killing vectors of FRW cosmologies one can also show that for a perturbed FRW universe, the long wavelength perturbations may be interpreted as conformal Killing motions of the FRW background. This allows to extend the consistency relations found in Refs. \cite{krls,ppls}  to the relativistic case \cite{creminellilss}.
We also show how one can extend  the
  conformal symmetry to the  full FRW bulk for accelerating cosmologies, following what is done in the AdS/CFT correspondence. In such a case the conformal symmetry is also an isometry of the FRW metric.
  
The paper is organized as follows.In section 2 we show that the accelerating FRW cosmologies on the future constant-time hypersurface exhibits three-dimensional conformal symmetry by embedding  their geometry in five-dimensional Minkowski space-time; 
we also    study an application of our findings, by showing  that
conformal invariance of scalar perturbations in an accelerating FRW universe can be obtained even away from de Sitter. 
In section 3 we offer an alternative
explanation which make use of conformal Killing vectors to show that  accelerating FRW cosmologies on the future constant-time hypersurface exhibits three-dimensional conformal symmetry and make use of these conformal Killing transformations to comment about the perturbed FRW cosmology in section 4. 
In section 5, we extend the boundary conformal symmetry to the bulk of accelerating FRW cosmologies, again presenting an explicit example
based on a scalar field. 
Finally, section 6 presents conclusions.

\section{Future boundary conformal symmetry of FRW accelerating cosmologies}
In this section we wish to make use of the embedding of the FRW space-time into five-dimensional
hyperboloids to show that FRW cosmology exhibits three-dimensional conformal symmetry on the future constant-time hypersurfaces. 
First, we start with the well-known case of de Sitter.

\subsection{Five-dimensional  hyperboloids and conformal symmetry of De Sitter}
It is well-known that in Euclidean three-dimensional space $\mathbb{R}^3$  conformal invariance is connected to the symmetry group SO(1,4) in the same way that 
conformal invariance in a four-dimensional Minkowski space-time is connected to the SO(2,4) group. As SO(1,4) is the 
isometry group of de Sitter space-time, one may conclude that 
during a  de Sitter stage of the evolution of the universe the isometry group acts  as conformal group  on $\mathbb{R}^3$ when the fluctuations
of a given quantum scalar field  are on super-Hubble scales. In such a  regime,  the SO(1,4) isometry
of the de Sitter background is realized as conformal symmetry of the flat $\mathbb{R}^3$ constant-time hypersurfaces sections  and   correlators of such fields are constrained by conformal invariance \cite{antoniadis,creminelli1,us1,us2}.
To show this explicitly, we can remind  some of the basic geometrical and algebraic properties of de Sitter space-time and group \cite{HE}.

The four-dimensional de Sitter space-time of constant radius $H^{-1}$ is described by the hyperboloid 
\be
\eta_{AB}X^AX^B=-X_0^2+X_i^2+X_5^2=\frac{1}{H^2} ~~~~( i=1,2,3),  \label{hyper}
\ee
embedded in  five-dimensional Minkowski space-time $\mathbb{M}^{1,4}$  with coordinates $X^A$ 
and flat metric  $\eta_{AB}=\rm{diag}\,(-1,1,1,1,1)$.
A particular parametrization of the de Sitter hyperboloid is provided by
\be
&&X^0=\frac{1}{2H}\left(H\tau-\frac{1}{H \tau}\right)-\frac{1}{2}\frac{\vx^2}{\tau},\nonumber \\
&&X^i=\frac{x^i}{H\tau},\nonumber \\
&&X^5=-\frac{1}{2H}\left(H\tau+\frac{1}{H \tau}\right)+\frac{1}{2}\frac{\vx^2}{\tau},  \label{poin}
\ee
which  satisfies Eq. (\ref{hyper}) as it can be easily checked. The de Sitter metric is the induced metric on the hyperboloid from 
the five-dimensional ambient Minkowski space-time
\be
{\rm d} s_5^2=\eta_{AB}\de X^A\de X^B.  \label{metric5}
\ee
For the particular parametrization (\ref{poin}), for example,  we find the familiar expression of the four-dimensional de Sitter metric with conformal time $\tau$ and scale factor $a(\tau)=-1/H\tau$
\be
 {\rm d}s_{\rm dS}^2=\frac{1}{H^2\tau^2}\left(-{\rm d}\tau^2+{\rm d} \vx^2\right).
\ee 
This metric is invariant under the infinitesimal coordinate transformations 
\be
&&\delta\t=\lambda \t, ~~~\delta_Dx^i=\lambda x^i, \nonumber \\
&&\delta\t=-2\t \, \vec{b}\cdot \vec{x}, ~~~\delta_Kx^i=-2 x^i\, 
(\vec{b}\cdot \vec{x})
+(\vx^2-\t^2)b^i, \label{iso1}
\ee
which together with rotations and translations form the de Sitter isometry group. 
Let us now focus on the limit $H\tau\ll 1$ or, equivalently, we look at future constant-time hypersurfaces  $\tau\to 0$.  For a generic scalar field this is equivalent at focussing onto those  fluctuations which are already outside the Hubble radius.
The parametrization (\ref{poin}) turns out then to be 
\be
&&X^0=-\frac{1}{2H^2\tau}-\frac{1}{2}\frac{\vx^2}{\tau},\nonumber \\
&&X^i=\frac{x^i}{H\tau}\nonumber, \\
&&X^5=-\frac{1}{2H^2\tau}+\frac{1}{2}\frac{\vx^2}{\tau} \label{par-hyp}
\ee
and we may easily check that the hyperboloid has been degenerated to the hypercone
\be
-X_0^2+X_i^2+X_5^2=0  \label{cone}.
\ee
We identify points $X^A\equiv \lambda X^A$ (which turns the cone (\ref{cone}) into a projective space).
 As a result, $\tau$ in the denominator
of the $X^A$ can be ignored due to projectivity condition. 
The conformal group SO(1,4) acts linearly on $X^A$, but it induces the (non-linear) conformal transformations $x_i\to x_i'$ with 
\be
&& x_i\to x_i'=a_i+M_i^{\,\,j}x_j, \label{c1}\\
&&x_i'=\lambda x_i, \label{c2} \label{c12}\\
&&x_i'=\frac{x_i+b_i \vx^2}{1+2\,\vec{b}\cdot\vx+b^2 \vx^2}. \label{c13}
\ee
One recognizes  the conformal group acting 
on Euclidean $\mathbb{R}^3$ with coordinates $x_i$. 
The infinitesimal form of (\ref{c12},\ref{c13}) i

For future use, we notice  that they 
 can be written also in terms of inversion
\be
{\rm  (inversion)}\times{\rm (translation)}\times{\rm (inversion)}, \,\,\,\,\,{\rm (inversion)}:
x_i\to x_i'=\frac{x_i}{\vx^2}, \label{inv}
\ee
We should also specify how $\t$ transforms 
under the conformal group. This can be determined either by the fact that $\t$ is a coordinate, or by the induced
metric on the cone, which turns out to be
\be
\de s^2=\frac{\de \vx^2}{H^2\tau^2}.
\ee
In both  ways one finds that  $\t$ transforms as
\be
&&\tau\to\tau'= \lambda \tau, \label{c21} \\
&&\tau\to \tau'=\frac{\tau}{\vec{x}^2}, \label{c22}
\ee
under dilations and inversions, respectively. 
As a result, the symmetry of the  constant future time hypersurfaces $\tau\to 0$
 is the three-dimensional conformal group generated by the transformations
(\ref{c1}), (\ref{c12}) and (\ref{c13}) augmented by the $\t$ transformations (\ref{c21}) and (\ref{c22}). 
Fluctuations of a generic scalar field (as well as tensor mode) must satisfy this symmetry on super-Hubble scales.

\subsection{Five-dimensional  hyperboloids and conformal symmetry of FRW accelerating cosmologies}
A generic FRW universe can similarly  be embedded in the five-dimensional Minkowski space-time $\mathbb{M}^{1,4}$  
with coordinates $X^A$ 
and flat metric  $\eta_{AB}=\rm{diag}\,(-1,1,1,1,1)$.
Indeed, it is easy to verify that the induced metric on the five-dimensional hyperboloid is now   determined by the parametrization (we present it only for the spatially flat FRW cosmology)
\begin{align}
X^0&=\frac{1}{2\tau_0} a(t)\left( \vx^2+\tau_0^2\right)+\frac{1}{\tau_0}\int^t \frac{\de t'}{2\dot{a}(t')}, \nonumber \\
 X^i&=a(t) x^i, \nonumber \\
X^5&=\frac{1}{2\t_0} a(t)\left( \vx^2-\t_0^2\right)+\frac{1}{\t_0}\int^t \frac{\de t'}{2\dot{a}(t')},
\end{align}
in $\mathbb{M}^{1,4}$, where dots indicate differentiation with respect to the cosmic time $t$ and $\t_0$ is a constant. 
The   metric (\ref{metric5}) becomes the familiar  FRW for a spatially flat sections
\be
{\rm d} s^2=-\de t^2+a^2(t)\de \vx^2,
\ee
where the  scale factor has the time dependence $a(t)\sim t^{2/3(1+w)}$, being  $w$  the equation of state of the 
fluid dominating the energy density of the universe at a given epoch.
It is also straightforward to determine the hypersurface which is embedded, as the de Sitter hyperboloid, in   $\mathbb{M}^{1,4}$. It is
\be
-X_0^2+X_1^2+X_2^2+X_3^3+X_5^2=(X_5+X_0) f(t), \label{hypersurface}
\ee
where 
\be
f(t)= \int^t \frac{\de t'}{\dot{a}(t')}=f_0 \, a^{2+3 w} \label{f}
\ee
and  $f_0$ is a constant. 
Note that   de Sitter is a particular case of this general embedding. 
In terms of the conformal time $\de\tau=\de t/a$ we have (for $w\neq -\frac{1}{3}$ and we set $a_0=a(\tau_0)=1$)
\be
a=(\tau/\tau_0)^{-q}, ~~~H\tau=-\frac{q}{a}, ~~~~q=-\frac{2}{1+3w},
\ee
where $H=\dot{a}/a$. Correspondingly,   we find that 
\be
f=-\frac{\t_0^2}{q}\left(-\frac{q}{H\tau}\right)^{2+3w}.
\ee
Note that the range of $q$ is such that 
\begin{eqnarray}
q\geq 1 &\Longleftrightarrow& -1\leq w< -\frac{1}{3},\nonumber \\
q\leq -\frac{1}{2} &\Longleftrightarrow& -\frac{1}{3}
<w\leq 1.
\end{eqnarray}
In conformal time the embedding equations turn out to be
\begin{align}
X^0&=-\frac{1}{2}\frac{\t_0}{q} \left\{\left(-\frac{q}{H\tau}\right)^{2+3w}+\frac{q}{H\tau}\right\}
-\frac{1}{2}\frac{q}{H\tau}\frac{\vx^2}{\t_0}, \nonumber \\
 X^i&=-q\frac{x^i}{H \tau},
\nonumber \\
X^5&=-\frac{1}{2} \frac{\t_0}{q}\left\{\left(-\frac{q}{H\tau}\right)^{2+3w}-\frac{q}{H\tau}\right\}+
\frac{1}{2}\frac{q}{H\tau}\frac{\vx^2}{\t_0}. 
\label{par-hyp2}
\end{align}
Now the similarity with the Eq. (\ref{par-hyp}) is obvious. In particular, it can be easily checked that 
the embedding coordinates (\ref{par-hyp2}) 
for $H\tau\ll1$ turn out to be
\begin{align}
X^0&=-\frac{1}{2} \frac{\t_0}{H\tau}-\frac{1}{2}\frac{q}{H\tau}\frac{\vx^2}{\t_0}, \nonumber \\
 X^i&=-q\frac{x^i}{H \tau},\,\,\,\,\,\,\,\,\,\,\,\,\,\,\,\,\,\, \,\,\,\,\,\,\,\,\,\,\,\,\,\,\,\,\,\,\,\,\,\,\,\,\,\,\, \,\,\,\,\,\,\,\,\,\,\,\,\,\,\,\,\,\,\,\,  (w<-1/3)\nonumber\\
X^5&=\frac{1}{2} \frac{\t_0}{H\tau}-\frac{1}{2}\frac{q}{H\tau}\frac{\vx^2}{\t_0},\label{par-hyp3}
\end{align}
and 
\begin{align}
X^0&=\frac{1}{2} \frac{\t_0}{q} \left(-\frac{q}{H\tau}\right)^{2+3w}-\frac{1}{2}\frac{q}{H\tau}\frac{\vx^2}{\t_0}, \nonumber \\
 X^i&=-q\frac{x^i}{H \tau},
\,\,\,\,\,\,\,\,\,\,\,\,\,\,\,\,\,\,\,\,\,\,\,\,\,\,\,\,\,\,\,\,\,\,\,\,\,\,\,\,\,\,\,\,\, \,\,\,\,\,\,\,\,\,\,\,\,\,\,\,\,\,\,\,\,\,\,\,\,\,\,\, (w>-1/3)
 \nonumber  \\
X^5&=\frac{1}{2} \frac{\t_0}{q} \left(-\frac{q}{H\tau}\right)^{2+3w}-\frac{1}{2}\frac{q}{H\tau}\frac{\vx^2}{\t_0}. \label{par-hyp4}
\end{align}
One can now  easily verify  that  for the embedding coordinates  (\ref{par-hyp3}), the hyperboloid 
degenerates to the cone
\be
-X_0^2+X_i^2+X_5^2=0.   \label{cone2}
\ee
This does not happen for the embedding coordinates (\ref{par-hyp4}).
Therefore, we conclude that FRW cosmology exhibits three-dimensional conformal symmetry on the future constant-time hypersurfaces  not only during a de Sitter $(w=-1)$ phase,  but
in fact for any accelerating $(w< -1/3)$ phase. 
The induced metric on the cone (\ref{cone2}) 
is
\be
\de s^2=\left(\frac{\tau_0}{\tau}\right)^{2q}\,\de \vx^2.
\ee
This metric is invariant under space translations and rotations 
\be
&& x_i\to x_i'=a_i+M_i^{\, j}x_j, \label{cc1}
\ee
as well as under dilations
\be
\tau\to\tau'= \lambda^z \tau, ~~~~x_i'=\lambda x_i, \label{cc2} 
\ee
and inversions
\be
\tau\to \tau'=\frac{\tau}{|\vec{x}|^{2z}}, ~~~~x_i\to x_i'=\frac{x_i}{\vec{x}^2}, \label{cc3}
\ee
where 
\be 
z=\frac{1}{q}. 
\ee
Alternatively, we can write the infinitesimal  dilations $D$ and special conformal 
transformations $K_i$ that leave the induced metric invariant, act as 
\be
&&\delta_D x_i=\lambda x_i, ~~~~\delta_D \t=z\lambda \,\t, ~~~~\delta_{D,K_i}\t_0=0,\nonumber\\
&& \delta_K\, x_i=-2 x_i(\vec{b}\cdot\vx)+b_i(-\tau^2+\vx^2), ~~~~~\delta_K\t=-2 q \t (\vec{b}\cdot\vx). \label{conf}
\ee
This is nothing but the Lifshitz conformal group acting 
on Euclidean $\mathbb{R}^3$ with coordinates $x_i$.  
 We conclude that the symmetry of the future constant-time hypersurfaces  $\tau\to 0$ is the three-dimensional conformal group generated by the transformations
(\ref{cc1}), (\ref{cc2}), and (\ref{cc3}).

\subsection{Conformal invariance of scalar perturbations in an accelerating FRW universe with boundary conformal symmetry}
Let us now show explicitly how the fluctuations of a scalar field can be rendered conformal invariant on super-Hubble scales thanks what we have learned in the previous section. 
Consider  the action of a free scalar field $\sigma(\tau,\vx)$  in a FRW background 
\be
S=\frac{1}{2}\int \de^4 x \sqrt{-g}\,
g^{\mu\nu}\,\partial_\mu\sigma\partial_\nu\sigma.
\label{act1}
\ee
In  a de Sitter background this theory is conformal invariant for invariant $\sigma$. However, 
 for an FRW metric of the form 
\be
\label{dfg}
\de s^2=\frac{-\de\tau^2+\de\vec{x}^2}{(\tau/\tau_0)^{2q}}, 
\ee
the action (\ref{act1}) is written as
\be
S=\frac{1}{2}\int 
\de^3x\de\tau\, \left(\frac{\tau_0}{\tau}\right)^{2q}\big{[}\sigma'^{2}-(\nabla\sigma)^{2}
\big{]},
\label{act22}
\ee
where the primes denote differentiation with respect to the conformal time $\tau$.
Clearly, the action  (\ref{act22}) is not  invariant under the transformations (\ref{cc2}) and (\ref{cc3}).
We can restore conformal invariance by 
considering instead the action  
\be
S=\frac{1}{2}\int \de^3x\de\tau\, 
I(\tau)\left(\frac{\tau_0}{\tau}\right)^{2q}\big{[}{\sigma'}^{2}-(\nabla\sigma)^{2/q}
\big{]},
\label{act21}
\ee
where $I(\tau)$ is an  appropriate function  of conformal time. It is easy to verify that $I(\tau)$ should be chosen as
\be
I(\tau)=\left(\frac{-\tau}{\t_0}\right)^{1-q}.
\label{II}
\ee
Such coupling can be easily obtained, for instance,  in a power-law inflationary model as follows \cite{I1}. In  power-law inflation \cite{pl} the 
inflaton potential is (we work with the Plankian mass set to unity)
\be
V(\phi)=V_0e^{-\sqrt{2s }\, \phi},  \label{potential}
\ee
where $s $ is some positive  coefficient.   The solution to the Einstein equations are
\be \label{sol0}
a(\tau)=\left(-\tau/\t_0\right)^{\frac{1}{s -1}}, 
~~~\phi=\frac{\sqrt{2 s }}{s -1}\ln\, (-\tau/\t_0)+\phi_0,
\ee
with the constant $\phi_0$ given by
\be
\phi_0=\frac{1}{\sqrt{2s}}\ln\left(\frac{V_0(s-1)^2}{2(s-3) \t_0^{2}}\right).
\ee
Therefore,   we may express $I(\t)$ as 
\be
I(\tau)=(-\tau/\t_0)^{1-q }=e^{\sqrt{\frac{s }{2}}\, (\phi-\phi_0)},
\ee
where $s=(q-1)/q$.
Thus, during a power-law inflationary phase, a non-minimally coupled  Lifshitz model to the inflaton with action  
\be
S=\frac{1}{2}\int \de^3x\de\tau \left(\frac{\t}{\t_0}\right)^{-3q+1}
\Big{\{}{\sigma'}^2-(\nabla\sigma)^{2/q}
\Big{\}}, ~~~~~
\label{act22}
\ee 
is conformal invariant if we assume that 
\be
\sigma(\tau,\vx)\to \sigma'(\tau,\vx),\,\,\,{\rm with}\,\,\, \sigma'(\tau',\vx')=\sigma(\tau,\vx). \label{inv}
\ee
To find the conformal dimension of the field $\sigma$ on super-Hubble  scales, we should look 
for time-dependent solutions for 
$\sigma_\tau=\sigma(\tau)$.
The scalar  $\sigma_\tau$ satisfies the equation
\be
\sigma_\tau''+\frac{1-3q}{\t}\sigma_\tau'
=0. \label{p1}
\ee
from where we find that 

\be
\sigma_\t= C_1+C_2 \t^{3q}  \label{sigmat}
\ee
It is easy to see, by using   

\be
q=-\frac{2}{1+3w}, ~~~~s =\frac{q-1}{q}=\frac{3}{2}(1+w),
\ee
that the exponent of $\tau$ above is always positive for acceleration
\be
q>0 ~~~\Longleftrightarrow~~~w<-\frac{1}{3}.
\ee
Therefore, the constant mode  dominates in Eq. (\ref{sigmat}) as $\t\to 0$,  giving rise to an exactly  scale-invariant power spectrum 
\be
\Big< \sigma_{\vk_1}\sigma_{\vk_2}\Big> \sim \frac{(2\pi)^3}{k^3}\delta^{(3)}(\vec{k}_1+\vec{k}_2).
\ee
Similarly, for deceleration $q<0$, the constant mode dominates again, now at the boundary $\t\to \infty$,
leading similarly to 
a scale-invariant spectrum.

We may also consider a massive scalar. In this case we should add a mass term to the action (\ref{act21}). However, 
a pure mass term will spoil conformal symmetry under (\ref{cc2}) and (\ref{cc3}) and one way to restore it is to consider instead 

\be
S=\frac{1}{2}\int \frac{\de^3x\de\tau}{(\t/\t_0)^{4q}}\, 
\left\{
I(\tau)\left(\frac{\tau_0}{\tau}\right)^{-2q}\big{[}{\sigma'}^{2}-(\nabla\sigma)^{2/q}
\big{]}-m^2 J(\t)\sigma^2\right\},
\label{act210}
\ee
where
\be
J(\t)=(-\t/\t_0)^{q-1}=e^{-\sqrt{\frac{s }{2}}\, (\phi-\phi_0)}.
\ee
In this case, the action (\ref{act210}) is explicitly written as
\be
S=\frac{1}{2}\int \de^3x\de\tau \left\{\left(\frac{\t}{\t_0}\right)^{-3q+1}
\Big{[}{\sigma'}^2-(\nabla\sigma)^{2/q}\Big{]}-m^2 \left(\frac{\t}{\t_0}\right)^{-3q-1}\sigma^2\right\}
, ~~~~~
\label{act220}
\ee 
and the corresponding field equation for a time-dependent field $\sigma_\t$ is 
\be
\sigma_\tau''+\frac{1-3q}{\t}\sigma_\tau'+\frac{m^2\t_0^2}{\t^2}\sigma_\t
=0. \label{p2}
\ee
The solution to this equation turns out to be
\be
\sigma_\t=\hat{C}_1 \t^{\Delta_-}+\hat{C}_2\t^{\Delta_+},
\ee
where 
\be
\Delta_\pm=\frac{3q}{2}\left(1\pm \sqrt{1-\frac{4m^2\t_0^2}{9q^2}}\right).
\ee
It is easy to verify that during acceleration ($q>0$) at  the $\t\to 0$ boundary, 
\be
\sigma_\t=\hat{C}_1 \t^{\Delta_-} \label{st}
\ee
dominates, which give rise a two-point function
\be
\Big< \sigma_{\vk_1}\sigma_{\vk_2}\Big> \sim \frac{(2\pi)^3}{k^{3-\Delta_-}}\delta^{(3)}(\vec{k}_1+\vec{k}_2). \label{2pt}
\ee
For $m\t_0\ll1$, we find that 
\be
\Delta_-\approx \frac{m^2\t_0^2}{3q}
\ee
and therefore  we get from (\ref{2pt}) an almost scale invariant red-shifted spectrum. 

Similarly, for deceleration, ($q<0$) at the $\t\to \infty$ boundary, the solution is again given by 
(\ref{st}) and the corresponding two-point function by (\ref{2pt}). However, in this case, for $m^2\t_0^2\ll1$ we have 
$\Delta_-<0$ and therefore the spectrum is almost scale invariant, but blue-shifted this time. 

Our findings imply that one cannot rule out power-law inflation \cite{pl} as a possible model for inflation and the cosmological perturbations. It is a common lore that power-law inflation is in a bad shape  in the light of the recent Planck data \cite{planck}: a potential
like (\ref{potential}) predicts a spectral index for the scalar perturbations $n_s=1 +2s/(s-1)$ and a tensor-to-scalar ratio $r=16s$; the current 95\% C.L. range on $n_s$ implies $10^{-2}\simlt s \simlt 1/40$, which in turn implies $0.16\simlt r\simlt 0.43$. This clashes with the 95\% C.L. Planck bound $r\simlt 0.12$. Our results show that a curvaton-like field coupled to the inflaton field via the action (\ref{act22}) acquires
a scale-invariant power spectrum on super-Hubble scales and the  tensor-to-scalar ratio can easily respect the Planck upper bound
by choosing $V_0$ appropriately small. 

\section{Conformal Killing Vectors and FRW accelerating cosmologies}
In this subsection, we wish to present an alternative derivation of the fact  
that any accelerating FRW universe enjoys the three-dimensional conformal group on the future constant-time hypersurface.  
Let us   consider the set of conformal motions on any spatially FRW background, {\it i.e.} the group of motions generated 
by vector fields ${\bf X}=X^\alpha\partial_\alpha$ satisfying the equation
\be
{\cal L}_{\bf X}g_{\alpha\beta}=\partial_\gamma g_{\alpha\beta} X^\gamma+g_{\alpha\gamma}\partial_\beta X^\gamma
+g_{\gamma\beta}\partial_\alpha X^\gamma=2\phi(x) g_{\alpha\beta}. \label{ckv}
\ee
The vectors ${\bf X}$ are called the  conformal Killing vectors (CKVs) and include as special cases Killing vectors $(\phi=0$),
dilation vectors  ($\partial_\alpha \phi=0$) or special conformal vectors  ($\partial_{\alpha\beta}^2\phi=0$). 

The set of CKV on a FRW background 
can be found, since FRW is conformal to Minkowski space-time, by using the fact that 
conformally related spaces have the same set of CKVs. Indeed, for two metrics $g_{\alpha\beta}$ and 
$\gamma_{\alpha\beta}=\rho^2g_{\alpha\beta}$, if $g_{\alpha\beta}$ satisfies  Eq. (\ref{ckv}), then the metric $\gamma_{\alpha\beta}$
satisfies \cite{ckv}
\be
{\cal L}_{\bf X}\gamma_{\alpha\beta}=2\,\psi\, \gamma_{\alpha\beta}, \label{ckv}
\ee
where 
\be
\psi={\bf X}\,\ln\, \rho+\phi.
\ee
The CKVs of four-dimensional Minkowski space-time are 
\be
&&{\bf P}_\alpha=\partial_\alpha, ~~~~~~~{\bf M}_{\alpha\beta}=x_\alpha\partial_\beta-x_\beta \partial_\alpha,\nonumber \\
&&{\bf D}=x^\alpha \partial_\alpha, ~~~~~{\bf K}_\alpha=2x_\alpha {\bf D}-x_\beta x^\beta {\bf P}_\alpha.
\ee
Of these ${\bf P}_\alpha$ (translations) and  ${\bf M}_{\alpha\beta}$ (rotations)  are Killing vectors ($\phi=0$) and generate isometries of Minkowski space-time
whereas, $D$ is the dilation ($\phi=1$) and ${\bf K}_\alpha$ generates the special conformal transformations
$(\phi=2x^a$). 
These vectors satisfy the SO(2,4) algebra defined by the non-zero commutation relations
\be
&&
[{\bf M}_{\alpha\beta},{\bf M}_{\gamma\delta}=\tau_{\alpha\delta}{\bf M}_{\beta\gamma}+
\tau_{\beta\gamma}{\bf M}_{\alpha\delta}-
\tau_{\alpha\gamma}{\bf M}_{\beta\delta}-
\tau_{\beta\delta}{\bf M}_{\alpha\gamma},\nonumber \\
&&[{\bf P}_\alpha,{\bf M}_{\beta\gamma}]=\tau_{\alpha\beta}{\bf P}_{\gamma}-\tau_{\alpha\gamma}{\bf P}_{\alpha}, ~~~
[{\bf K}_\alpha,{\bf M}_{\beta\gamma}]=\tau_{\alpha\beta}{\bf K}_{\gamma}-\tau_{\alpha\gamma}{\bf K}_{\alpha},\nonumber \\
&&[{\bf P}_\alpha,{\bf K}_{\beta}]=2(\tau_{\alpha\beta}{\bf D}-2{\bf M}_{\alpha\beta}), 
~~~[{\bf D},{\bf K}_\alpha]={\bf K}_\alpha, ~~~[{\bf D},{\bf P}_\alpha]=-{\bf P}_\alpha.
\ee
Let us now consider the  general 
FRW metric in conformal time
\be
\de s^2=\left(\frac{\t_0}{\t}\right)^{2q}\left(-\de\tau^2+\de\vec{x}^2\right), \label{frw1}
\ee
which also includes as special cases Minkowski ($q=0$) and de Sitter ($q=1$). The CKVs and the corresponding conformal
factor ($\phi$ for Minkowski and $\psi$ for de Sitter or generic FRW) are given in the following table
\be
\centering
\begin{array}{l c c c}
\hline\hline
 & \mbox{  Minkowski} & \mbox{ de Sitter} & \mbox{ generic FRW} \\ [0.5ex]
\hline
{\bf P}_0 & 0 & -\frac{1}{\tau} & -q\frac{1}{\tau} \\
{\bf P}_i & 0 & 0 & 0 \\
{\bf M}_{ij} & 0 & 0 & 0 \\
{\bf M}_{0i} & 0 & \frac{x_i}{\tau} & q \frac{x_i}{\tau}\\
{\bf D}  & 1 & 0 & q-1 \\ 
{\bf K}_0 & 2\tau & \frac{\tau^2+\vec{x}^2}{\tau} & \frac{(2q-1)\tau^2+\vec{x}^2}{\tau} \\
{\bf K}_i & 2x_i & 0 & 2(q-1)x_i \\ [1ex]
\hline
\end{array}
\label{table:nonlin}
\label{table}
\ee
It is easy to see that 
the metric has a boundary at $\tau\to 0$ for $q>0$ and a boundary at $\tau\to \infty$ for
$q<0$. In other words, accelerating
backgrounds process a boundary at  $\tau\to 0$ whereas for decelarating ones the boundary is at $\tau\to \infty$. 
At the  boundary,
the FRW CKVs can be projected into tangential and normal to the boundary. It is easy to verify that the tangential vectors 
turn out to be
\be
&&\hat{{\bf P}}_i=\partial_i, ~~~~~~~\hat{{\bf M}}_{ij}=x_i\partial_j-x_i \partial_j,\nonumber \\
&&\hat {\bf D}=x^i \partial_i, ~~~~~\hat {\bf K}_i=2x_i \hat {\bf D}-x_i x^j \hat {\bf P}_i, ~~~(i,j=1,2,3).
\ee
We conclude that  for an accelerating universe the $\tau\to 0$  boundary process ten CKVs which form an SO(1,4) algebra, 
precisely the three-dimensional conformal algebra.
Only for de Sitter space $q=1$, however, these ten  CKVs are actually isometries,  whereas for all the other cases
($q<1$) only $\hat{{\bf P}}_i$ and $\hat{{\bf M}}_{ij}$ generate isometries.    
Therefore, if a theory is invariant under the generators $\hat{{\bf P}}_i,\hat{{\bf M}}_{ij},\hat {\bf D}$ and $\hat {\bf K}_i$
in the bulk of an accelerating FRW cosmology, it will be automatically SO(1,4) invariant, {\it i.e.}  conformal invariant at the boundary $\tau\to 0$.

\section{Perturbed FRW universe and conformal symmetries}
As an application of what we have described in the previous section, let us consider now a perturbed FRW metric and write the corresponding metric in, say, the Newtonian  gauge

\be
\de s^2 =\left(\frac{\tau_0}{\tau}\right)^{2q}\left[-(1+2\Phi)\de \tau^2+(1-2\Phi)\de \vx^2\right],
\label{per}
\ee
where $\Phi(\tau,\vx)$ is the gravitational potential. 
We restrict ourselves to long wavelength part 
of the perturbations, indicated by $\Phi_L$ , which are still in the linear regime. More in particular, 
we make the following assumptions: $i$) the long wavelength gravitational potential does not evolve in 
time and $ii$) we consider such perturbations only up to first order both in $\Phi_L$  and its gradient 
$\partial_i\Phi_L$. The first condition is automatically attained during the matter-dominated (MD) period 
while in the radiation-dominated (RD) period it is valid thanks to the second condition, 
as $\dot\Phi_L={\cal O}(c_s^2\tau^2\nabla^2\Phi_L)$, where $c_s\simeq 1/\sqrt{3}$ is the sound speed of the 
perturbations (notice that during the matter-dominated period $c_s\simeq 0$ and therefore $\Phi_L$ is constant 
even on very small scales). Notice that the gravitational potential $\Phi_L$ is related to the comoving curvature 
perturbation $\zeta_L$ whose value is set by inflation, 
$\Phi_L({\rm MD})=9/10\Phi_L({\rm RD})=3/5\zeta_L$. Notice also that the following considerations hold 
also during inflation,  {\it i.e.} during a period of quasi de Sitter expansion.

Our goal is  to show that the long wavelength perturbations (in the Newtonian gauge)  are just (projected) conformal Killing motions of the FRW background. Let us first perform a direct computation and  write the metric (\ref{per}) as 

 \be
\de s^2 =\left(\frac{\tau_0}{\tau}\right)^{2q}(1+2\Phi_L)\left[-\de \tau^2+(1-4\Phi_L)\de \vx^2\right],
\label{per1}
\ee
and perform an infinitesimal  dilation transformation followed by an infinitesimal special conformal 
transformation on the coordinates $x^i$

\be
 y^i=x^i(1+\lambda)-2(\vx\cdot\vec{b})x^i+b^i\vx^2.
\ee
Since this is a three-dimensional conformal transformation in flat $\mathbb{R}^3$, it implies that

\be
\de \vec{y}^2= \left(1+2\lambda-4\vec{b}\cdot \vx\right)\,\de \vx^2.
\ee
Expanding the long wavelength part of the gravitational potential around an arbitrary point $\vx_0$ (which we choose to set as the origin of the coordinates),

\be
\Phi_L(\vx)\simeq \Phi_L(\vec 0)+\partial_i\Phi_L(\vec 0)\,x^i +{\cal O}(\partial_i\partial_j\Phi_L(\vec 0)),
\label{fl}
\ee
and choosing 

\be
\lambda=-2\Phi_L(\vec 0),\,\,\,\,\,\,b^i=\partial^i\Phi_L(\vec 0),
\ee
we can recast the metric (\ref{per1}) in the form 

\be
\de s^2=\t^2\left(\frac{\t_0}{\t}\right)^{2q}(1+2\Phi_L)
\left(\frac{-\de\tau^2+\de\vec{y}^2}{\t^2}\right),\,\,\,\,\, \,\,
\label{per3}
\ee
which is conformal to the de Sitter metric. Since following infinitesimal transformation

\be
 z^i=y^i(1+\alpha)-2(\vec{y}\cdot\vec{a})y^i+a^i(-\t^2+\vec{y}^2),\,\,\,\,\eta=\tau(1+\alpha-2\vec{y}\cdot\vec{a}), 
 \label{tra1}
\ee
is an isometry of the de Sitter metric, and choosing

\be
\alpha=\frac{1}{1-q}\Phi_L(\vec 0),\,\,\,\,\,\,a^i=\frac{1}{2(q-1)}\partial^i\Phi_L(\vec 0),
\ee
we can recast the metric (\ref{per3}) into the form

\be
\de s^2=\eta^2\left(\frac{\t_0}{\eta}\right)^{2q}
\left(\frac{-\de\eta^2+\de\vec{z}^2}{\eta^2}\right),\,\,\,\,\, \,\,s=1-\frac{1}{q},
\label{per3}
\ee
which is the background FRW metric.
In terms of the original coordinates in (\ref{per}), the transformation (\ref{tra1}) is written as 

\begin{align}
z^i&=x^i\left(1+\frac{2q-1}{1-q}\Phi_L(\vec{0})\right)+\frac{1}{2(q-1)}\Big{\{}-\t^2+(2q-1)\vec{x}^2\Big{\}}\partial^i
\Phi_L(\vec{0})
+\frac{1-2q}{q-1}\Big{(}\vec{x}\cdot \vec{\nabla}\Phi_L(\vec{0})\Big{)}x^i \nonumber \\
&=x^i\left(1+\frac{2q-1}{1-q}\Phi_L(\vec{x})\right)+\frac{1}{2(q-1)}\Big{\{}-\t^2+(2q-1)\vec{x}^2\Big{\}}\partial^i
\Phi_L(\vec{x}), \nonumber \\
\eta&=\tau\left(1+\frac{1}{1-q}\Phi(\vec{0})+\frac{1}{1-q}\vec{x}\cdot \nabla\Phi_L(\vec{0})\right)=
\tau\left(1+\frac{1}{1-q}\Phi_L(\vec{x})\right).  \label{per4}
\end{align}
Note that during single-field inflation, where the comoving curvature perturbation is  generated by an inflaton field $\phi(\tau,\vx)$ with potential $V(\phi)$ and one has a period of quasi de Sitter, the parameter 
$q$ is given by $q=(1+\epsilon)$, being $\epsilon=(1-{\cal  H}'/{\cal H}^2)$ one of the slow-roll parameters. 
However, the change of coordinates is not singular in the de Sitter limit of tiny $\epsilon$ as, 
in the Newtonian gauge, $\Phi_L(\vec{x})=\epsilon {\cal H}\delta\phi(\vec{x})/\phi'$.
It is easy to see that the transformations (\ref{per4}) in this case reduces to (\ref{iso1}), {\it i.e.} they are just 
an isometry of the background and therefore leave the metric invariant. 

For a matter dominated era $q=-2$,  so that the conformal transformation that eliminates (or, equivalently, 
generates)
long wavelength comoving perturbations, is 

\begin{align}
z^i&=x^i\left(1-\frac{5}{3}\Phi_L(\vec{x})\right)+
\left(\frac{1}{6}\t^2+\frac{5}{6}\vec{x}^2\right)\partial^i
\Phi_L(\vec{x}), \nonumber
 \\
\eta&=\tau\left(1+\frac{1}{3}\Phi(\vec{x})
\right). \label{contra}
\end{align}
These relativistic transformations generalize the non-relativistic ones found in Refs. \cite{krls,ppls} and have been recently derived in Ref.  \cite{creminellilss}. They allow to
derive  relativistic consistency relations involving the soft limit of the $(n + 1)$-correlator 
functions of matter  overdensities and the primordial gravitational potential \cite{creminellilss}. 
These consistency relations, as pointed out in Refs.  \cite{krls,creminellilss},  are valid not only for dark matter, but also for halos and galaxy number density,  and might be used to test theories of modified gravity.
 
We wish to  show now that the 
long wavelength comoving curvature perturbations in the Newtonian gauge are indeed generated by CKVs. 
The reason behind  this is that 
 long wavelength  perturbations are linear in the spatial coordinates as can be seen from Eq. (\ref{fl}).
As a result, they can be generated by  dilations and 
special conformal transformations. 
In particular, it is straightforward to verify that  

\be
{\cal{L}}_{{\bf{\xi}}_1} g_{ij}= (1+2\lambda -4\vec{b}\cdot\vx)g_{ij}, ~~~
{\cal{L}}_{{\bf{\xi}}_1} g_{00}= 0,~~~{\cal{L}}_{{\bf{\xi}}_1} g_{0i}= 0,
\ee
where ${\bf \xi}_1$ is the vector

\be
{\bf \xi}_1=(1+2\lambda)x^i\partial_i+\Big{\{}b^i\vec{x}^2-2(\vec{b}\cdot\vec{x})\Big{\}}\partial_i.
\ee
This means that  ${\bf \xi}_1$ is nothing else than a linear combinations of the CKVs 
tangent to the $\t={\rm const.}$ hypersurfaces. Indeed, ${\bf \xi}_1$ can be written as 
\be
{\bf \xi}_1=(1+2\lambda) {\Pi(\bf{D}}) -b^i{\Pi({\bf K}}_i),
\ee
where ${\Pi(\bf{D}}),{\Pi({\bf K}}_i)$ are the projections of the CKV along the $\t={\rm const.}$ hypersurfaces. 
Therefore, ${\bf \xi}_1$ generates the factor $(1-4\Phi_L)$ in front of the spatial part of the metric (\ref{per1}). Then 
it is easy to verify by a simple inspection of the table (\ref{table}), that the conformal factor $(1+2\Phi_L)$
is generated simply by the CKV ${\bf \xi}_2$ 

\be
{\bf \xi}_2=\frac{s-1}{2s}\, (1-\lambda) {\bf D}+\frac{s-1}{2s}\,  b^i {\bf{K}}_i,
\ee
since we have 

\be
{\cal{L}}_{{\bf{\xi}}_2} g_{\mu\nu}=(1-\lambda+2 \vec{b}\cdot\vx)g_{\mu\nu}.
\ee
Therefore, long wavelength comoving curvature perturbations  of the FRW metric 
are just conformal Killing motions of the FRW background.  

For the sake of completeness, let us offer another, direct derivation of the transformation
(\ref{contra}). Consider the metric (\ref{dfg})
and make the change of coordinates $
\t\to \t+\eta(\t,\vx)$ and $x_i\to x^i+\xi_i(\t,\vx)$.
Under this transformation, the metric  (\ref{dfg})  changes to 
\be
\de s^{2}=\left(\frac{\t_0}{\t}\right)^{2q}\left\{-\Big(1-2q\frac{\eta}{\t}
+\d_\t\eta\Big) \de\t^2+\Big(1-2q\frac{\eta}{\t}\Big) \de\vec{x}^2+2\d_j\xi_i \de x^i\de x^j+2
\Big( \d_\t\xi_i-\d_i\eta\Big)\de\t \de x^i\right\}. \label{bigm}
\ee
This is of the form (\ref{per}) if the following conditions are satisfied
\be
&& \d_i\eta=\d_\t \xi_i, \label{q1}\\
&&\d_\t\eta-q\frac{\eta}{\t}=\Phi_L, \label{q2}\\
&&\d_i\xi_j+\d_j\xi_i=2\delta_{ij}\left(q\frac{\eta}{\t}-\Phi_L\right).\label{q3}
\ee
The solution of Eq. (\ref{q1}) is 
\be
\eta=\frac{1}{1-q}\t \, \Phi_L+\t^q c(\vx), \label{q4}
\ee
so that  Eq. (\ref{q2}) gives
\be
\xi_i=\frac{\t^2}{2(1-q)}\d_i\Phi_L+\frac{\t^{q+1}}{q+1}\d_i c(\vx)+f_i(\vx).
\ee
Finally the last equation (\ref{q3}) turns out to be
\be
2\frac{\t^{q+1}}{q+1}\d_i\d_j c+ \d_if_j+\d_jf_i=2\frac{2q-1}{1-q}\Phi_L \delta_{ij}+2q\t^{q-1}c\delta_{ij}.
\ee
Therefore, 
\be
c=0, ~~~~~\d_if_j+\d_jf_i=2\frac{2q-1}{1-q}\Phi_i \delta_{ij} \label{q5}
\ee
and thus $f_i$ are just conformal Killing vectors.
The solution to Eq. (\ref{q5}) as we have seen in section 3 is 
\be
f_i=2a^jx_j x_i-\vx^2 a_i+\lambda x_i,
\ee
where, for the present case,
\be
a_i=\frac{2q-1}{2(1-q)}\d_i \Phi_L(\vec{0}), ~~~~\lambda=\frac{2q-1}{1-q}\Phi_L(\vec{0}).
\ee
Then, it is straightforward to verify that 
\be
\eta=\frac{1}{1-q}\t \, \Phi_L(\vec{0}), ~~~~\xi_i=\frac{2q-1}{1-q}x_i\Phi_L(\vec{0})+\frac{1}{2(1-q)}\left\{
\t^2-\big(2q-1\big)x^2\right\} \d_i \Phi(\vec{0}),
\ee
so that we get the result (\ref{per4}).
Note that for a de Sitter space, $q=1$ and the transformation is singular. The reason is that the integration 
of Eq. (\ref{q1}) for the $q=1$ case gives instead 
\be
\eta=\Phi_L(\vec{0})\, \t \log\t+\t c(\vx), \label{q4}
\ee
which leads to 
\be
\xi_i=\d_i \Phi_L(\vec{0})\left(\frac{1}{2}\t^2\log\t-\frac{\t^2}{4}\right)+\frac{\t^2}{2}\d_ic(\vx)+g_i(\vx).
\ee
Eq. (\ref{q3}) is then written as 
\be
2\frac{\t^2}{2}\d_i\d_jc+\d_ig_j+\d_jg_i=2\left\{\Phi_L(\vec{0})(\log\t-1)+c\right\}\delta_{ij},
\ee
which does not have a solution. Long wavelength  perturbations in de Sitter cannot be generated in this way. 
This is consistent with the fact that there are no scalar perturbations in de Sitter background. This is also consistent with the fact that  conformal Killing motions in de Sitter space-time are 
just isometries and therefore long wavelength  perturbations cannot be generated.



\section{Bulk conformal symmetry of FRW accelerating cosmologies}
In the previous section we have seen that the future boundary of any FRW accelerating cosmology enjoys conformal invariance.  On the other hand, it is well known that in the AdS/CFT correspondence   as well as in any holographic theory, the symmetries
of the boundary theory correspond to isometries of the bulk. Translating this in the present context, one should expect
that since the boundary theory enjoys conformal invariance, the isometries of the bulk of an accelerating FRW is 
 the full conformal group. However, the  metric (\ref{frw1})
is clearly not invariant under the three-dimensional  conformal group. We may write it  as
\be
\label{dsc}
\de s^2=\left(\frac{\t^s}{\t_0}\right)^\frac{2}{s-1}
\left(\frac{-\de\tau^2+\de\vec{x}^2}{\t^2}\right)=\Omega^2\, \de s_{\rm dS}^2, \,\,\,\,\,\,s=1-\frac{1}{q},
\ee
that is  in a conformally de Sitter form where $\de s^2_{\rm dS}$ is the de Sitter metric and 
\be
\Omega^2=\left(\frac{\t^s}{\t_0}\right)^\frac{2}{s-1}. \label{Omega}
\ee
Then, since de Sitter space-time has  the three-dimensional conformal group SO(1,4) as isometry group, the failure of the FRW metric to 
have the same symmetries is due to the conformal factor $\Omega$. Therefore, to  implement SO(1,4) invariance 
for an FRW background,  we have to demand that

\be
\delta_D\Omega=0, ~~~\delta_{K_i}\Omega=0, \label{omega}
\ee
under infinitesimal  dilations $D$ and special conformal transformations $K_i$. 
One possibility is the following. Consider the action
\be
S=\int \de^4x\, \sqrt{-g}\left(\frac{1}{2}\varphi^2 R+3\partial_\mu\varphi\partial^\mu\varphi-
\frac{1}{2}\varphi^2\partial_\mu\phi\partial^\mu\phi -\varphi^4V_0e^{-\sqrt{2s }\, \phi}\right),
\ee
where we have introduced an additional field (compensator or Stuckelberg field)  $\varphi$.  
Note that the kinetic term for the field $\varphi$ has opposite sign and seems to describe a ghost \cite{kallosh}. 
However, $\varphi$ 
can be fixed at will as the theory is invariant under
conformal rescalings
\be
g_{\mu\nu}(x)\to e^{2\omega(x)}g_{\mu\nu}(x), ~~~~\varphi(x)\to e^{-\omega(x)}\varphi(x) 
, ~~~\phi(x)\to \phi(x).
\label{conf1} 
\ee
Indeed, the action (\ref{conf1}) can be written in terms of the invariant metric $\varphi^2 g_{\mu\nu}$ 
and a solution is provided by 
\be
\de s^2=  g_{\mu\nu}\de x^\mu \de x^\nu
=\varphi^{-2}\left(\frac{\tau}{\t_0}\right)^{\frac{2}{s -1}}\left(-\de\t^2+\de\vec{x}^2\right)
, ~~~\phi=\sqrt{\frac{2}{s} }\frac{1}{s -1}\ln\, \left(\frac{\tau^s}{\t_0}\right)+\widetilde{\phi}_0, \label{sol2}
\ee
where
\be
\widetilde \phi _0=\frac{1}{\sqrt{2s}}\ln\left(\frac{V_0(s-1)^2}{2(s-3) }\right).
\ee
In fact, we may express (\ref{sol2}) as
\be
\de s^2
=\varphi^{-2}e^{\sqrt{2s} (\phi-\widetilde \phi _0)}\left(\frac{-\de\t^2+\de\vec{x}^2}{\t^2}\right)
, ~~~\phi=\sqrt{\frac{2}{s} }\frac{1}{s -1}\ln\, \left(\frac{\tau^s}{\t_0}\right)+\widetilde{\phi}_0. \label{sol21}
\ee
Fixing the value of $\varphi$ corresponds to choosing the conformal frame and removes the freedom (\ref{conf1}).
Interesting gauges are the $\varphi=1$ gauge
which is just the Einstein frame and gives the solution (\ref{sol0}). Another gauge is
\be
\varphi=e^{\sqrt{\frac{s}{2}} (\phi-\widetilde \phi _0)}, ~~~\de s^2=\frac{-\de\t^2+\de\vec{x}^2}{\t^2}, 
~~~\phi=\sqrt{\frac{2}{s} }\frac{1}{s -1}\ln\, \left(\frac{\tau^s}{\t_0}\right)+\widetilde{\phi}_0, \label{sol3}
\ee
which gives a de Sitter background metric.
 The metric in Eq. (\ref{sol2}) can be written as 
\be
\de s^2= \widetilde{\Omega}^2\left(\frac{-\de\t^2+\de\vec{x}^2}{\t^2}\right),\,\,\,\,\,\,\,\,
\widetilde{\Omega}^2=\varphi^2 \frac{\t_0^{2q}}{\t^{2(q-1)}}=\varphi^2\Omega^2,
\ee
and $\widetilde{\Omega}^2$   reduces to (\ref{Omega}) in the $\varphi=1$ gauge. Now, under a conformal transformation
$\widetilde \Omega$ satisfies 
\be
\delta_D\widetilde \Omega=0, ~~~\delta_{K_i}\widetilde \Omega=0. \label{omega}
\ee
 In other words, the metric (\ref{dsc}) does not look
invariant under conformal transformations  simply because is written in the particular conformal frame
$\varphi=1$, the Einstein frame.  In fact, under a conformal transformation, one should transform $\varphi$ as 
well before setting $\varphi=1$, thus rendering manifest the bulk conformal symmetry of FRW accelerating cosmologies. 
Notice that the transformation of the conformon field $\varphi$ may be traded with the transformation of  
the dimensional quantity $\tau_0$. This is not strange  as $\tau_0$ may be thought as a scalar field being  
directly linked to the initial value $\phi_0$ of the vacuum expectation value of the inflaton field,  
see Eq. (\ref{sol2}).

\subsection{Conformal invariance of scalar perturbations in an accelerating FRW universe with bulk conformal symmetry}
Let us now make use of what we have just learnt and try to construct a conformally-invariant action 
for a scalar field which will deliver a flat-invariant spectrum on super-Hubble scales. 

Let us consider now a scalar field $\sigma(\t,\vx)$, the dynamics of which is determined by the action
\be
S=-\frac{1}{2}\int \de^4 x \sqrt{-g}\, \varphi^2\, I^2(\phi)\,
g^{\mu\nu}\,\partial_\mu\sigma\partial_\nu\sigma, \label{sim}
\ee
where $I(\phi)$ represents a coupling to the inflaton. Note that (\ref{sim}) is 
 invariant under the transformation  (\ref{conf1}).  
Using (\ref{sol21}), the action  (\ref{sim}) is independent of the conformal frame as the $\varphi$ factors drop out
and we get
\be
S=-\frac{1}{2}\int \de^3x\de\tau\,   
I^2(\phi) \, e^{-\sqrt{2s} (\phi-\widetilde \phi _0)}\frac{\t_0^2}{\t^2}\, \Big{\{}{\sigma'}^{2}-(\nabla\sigma)^{2}\Big{\}}
\label{act21}
\ee
Then, for 
%
\be
I(\phi)=e^{-\sqrt{\frac{s}{2}} (\phi-\widetilde \phi _0)}, \label{I1}
\ee
the action (\ref{sim}) is explicitly written as 
\be
S=\frac{1}{2}\int \de^3x\de\tau  
\frac{\t_0^2}{\t^2}\Big{\{}{\sigma'}^{2}-(\nabla\sigma)^{2}\Big{\}}
\label{act21}
\ee
Clearly, (\ref{act21}) is conformal invariant, 
%
if we assume that 
\be
\sigma(\tau,\vx)\to \sigma'(\tau,\vx),\,\,\,{\rm with}\,\,\, \sigma'(\tau',\vx')=\sigma(\tau,\vx). \label{inv}
\ee
To find the conformal dimension of the field $\sigma$ on super-Hubble  scales, 
we should look as usual for time-dependent solutions for 
$\sigma_\tau=\sigma(\tau)$ to the equations of motion 
\be
\sigma_\t''-\frac{2}{\t}\sigma_\t'=0.
\ee
Then we find that 
\be
\sigma_\t=\sigma_0+\sigma_1\t^3,
\ee
so that on super-Hubble scales we will have the constant mode
\be
\sigma(\t,\vx)=\sigma_0(\vx).
\ee 
The conformal dimension of $\sigma_0(\vx)$ is then $\Delta_{\sigma_0}=0$ and therefore, we will have an exact scale-invariant spectrum
\be
\Big<\sigma_{\vk_1}\sigma_{\vk_2}\Big>\sim \frac{(2\pi)^3}{k_1^3}\delta^{(3)}(\vec{k}_1+\vec{k}_2).
\ee
The corresponding conformal weight of the initial field $\sigma(\t,\vx)$ can be easily deduced from the transformations of the constant-time
$\tau_0$.

Note that, the above findings are valid for any FRW geometry, including decelerating one. However, in the latter 
case, as the boundary is at $\t\to \infty$, the mode that dominates is the $\sigma_1$ which has conformal dimension $
\Delta_{\sigma_1}=3$.
As a result, the two-point correlator is given in this case by
\be
\Big<\sigma_{\vk_1}\sigma_{\vk_2}\Big>\sim (2\pi)^3\,\,k_1^3\, \delta^{(3)}(\vec{k}_1+\vec{k}_2) ~~~~~~
\mbox{(deceleration)}.
\ee
Finally, we may also consider massive scalars described by 
\be
S=-\frac{1}{2}\int \de^4 x \sqrt{-g}\, \varphi^2\, I^2(\phi)\,
\left(g^{\mu\nu}\,\partial_\mu\sigma\partial_\nu\sigma+m^2 \varphi^2\sigma^2\right). \label{simm}
\ee
This action is again invariant under the transformation (\ref{conf1}). By using 
(\ref{sol21}), the action  (\ref{sim}) is also independent of the conformal frame  ($\varphi$ factors again drop out)
and it is explicitly written as

\be
S=\frac{1}{2}\int \de^3x\de\tau\,   
I^2(\phi) \, e^{-\sqrt{2s} (\phi-\widetilde \phi _0)}\, 
\left\{\frac{\t_0^2}{\t^2}\Big{(}{\sigma'}^{2}-(\nabla\sigma)^{2}\Big{)}-m^2\sigma^2\right\}.
\label{act212}
\ee
Thus, (\ref{I1}) reduces (\ref{act212}) to 

\be
S=\frac{1}{2}\int \de^3x\de\tau\, 
\left\{\frac{\t_0^2}{\t^2}\Big{(}{\sigma'}^{2}-(\nabla\sigma)^{2}\Big{)}-m^2\sigma^2\right\},
\label{act213}
\ee
which is the standard action of a massive scalar on a de Sitter background. The two-point correlator is in this case

\be
\Big<\sigma_{\vk_1}\sigma_{\vk_2}\Big>\sim \frac{(2\pi)^3}{k_1^{3-2\Delta}}\delta^{(3)}(\vec{k}_1+\vec{k}_2)
, ~~~~\Delta=\frac{3}{2}\left(1-\sqrt{1-\frac{4}{9}m^2\t_0^2}\right).
\ee
We stress that the above results are frame independent  as the conformon field disappears from the action.

\section{Conclusions}
In this paper we have shown that any   accelerating FRW cosmology
enjoys a three-dimensional Lifshitz  conformal symmetry on the future constant-time hypersurface. It has been recently pointed out that
flat FRW cosmologies and higher-dimensional hyperscaling-violating geometries can be connected by analytic continuation 
\cite{hyper}. It would be interesting to investigate this connection further in the light of our results. 
We have also shown that the boundary conformal symmetry can be extended to the bulk of FRW accelerating cosmologies. This is reminiscent of the so-called  generalized conformal symmetry which was proposed as an extension of the conformal symmetry of
(boundary) D3-branes to general (bulk) D$p$-brane systems   \cite{jevicki}.

Our results imply that   one can construct a theory of a free scalar field
in an accelerating FRW cosmology whose perturbations will be scale-invariant and, in fact, even conformal invariant on super-Hubble scales. 
On general grounds this means that measuring a scale-invariant power spectrum for the cosmological perturbation does not  imply that the universe went necessarily through a de Sitter stage. Indeed, it suffices to have a curvaton-like field  \cite{curvaton1,curvaton2,curvaton3} appropriately coupled 
to the any accelerating FRW cosmology and its perturbations will be scale-invariant. Furthermore, its  three-point correlator will respect
the full three-dimensional conformal  symmetry on super-Hubble scales and therefore it will be enhanced in the squeezed configuration \cite{creminelli1,us1}.

\section*{Acknowledgments}
 We thank P.~Creminelli, J.~Norena, M.~Simonovic and F. Vernizzi for useful correspondence. A.R. is supported by the Swiss National
Science Foundation (SNSF), project `The non-Gaussian Universe" (project number: 200021140236).
This research was implemented
under the ARISTEIA Action of the Operational Programme Education and Lifelong Learning
and is co-funded by the European Social Fund (ESF) and National Resources. It is partially
supported by European Union’s Seventh Framework Programme (FP7/2007-2013) under REA
grant agreement n. 329083.


\end{document}